\documentclass[12pt]{article}
\usepackage[reqno]{amsmath}
\usepackage{bbm}
\usepackage{epsfig}
\usepackage{array}
\usepackage{float}
\usepackage{dsfont}
\usepackage{amstext}

\usepackage{a4}

\usepackage{a4wide}
\usepackage{wasysym}

\def\be{\begin{equation}}
\def\ee{\end{equation}}
\def\gs{\mathrel{
   \rlap{\raise 0.511ex \hbox{$>$}}{\lower 0.511ex \hbox{$\sim$}}}}
\def\ls{\mathrel{
   \rlap{\raise 0.511ex \hbox{$<$}}{\lower 0.511ex \hbox{$\sim$}}}}

\newcommand{\ba}{\begin{array}{c}}
\newcommand{\baz}{\begin{array}{cc}}
\newcommand{\bad}{\begin{array}{ccc}}
\newcommand{\bea}{\begin{equation} \begin{array}{c}}
\newcommand{\eea}{ \end{array} \end{equation}}
\newcommand{\ea}{\end{array}}

\newcommand{\dms}{\mbox{$\Delta m^2_{\odot}$}}
\newcommand{\dma}{\mbox{$\Delta m^2_{\rm A}$}}
\newcommand{\meff}{\mbox{$\langle m \rangle$}}

\begin{document}

\title{\vspace{-2cm}
\hfill {\small TUM--HEP--596/05}\\
\vspace{-0.3cm}
\hfill {\small hep--ph/0507143} 
\vskip 0.2cm
\bf 
Deviations from Tribimaximal Neutrino Mixing}
\author{
Florian Plentinger\thanks{email: \tt florian$\_$plentinger@ph.tum.de}~\mbox{ }~~and~~
Werner Rodejohann\thanks{email: \tt werner$\_$rodejohann@ph.tum.de} 
\\\\
{\normalsize \it Physik--Department, Technische Universit\"at M\"unchen,}\\
{\normalsize \it  James--Franck--Strasse, D--85748 Garching, Germany}
}
\date{}
\maketitle
\thispagestyle{empty}
\vspace{-0.8cm}
\begin{abstract}
\noindent 
Current neutrino data are consistent with the so--called tribimaximal 
mixing scenario, which predicts $\sin^2 \theta_{12}=1/3$, zero $U_{e3}$ and 
maximal $\theta_{23}$. 
This implies a special form of the neutrino mass matrix. 
Introducing small breaking terms in this 
mass matrix generates deviations from the tribimaximal scheme and leads 
to testable correlations between the parameters. They depend on where 
the perturbation is located in the mass matrix. 
A special case of such perturbations are radiative corrections. 
Alternative deviations from tribimaximal mixing may stem from contributions 
of the charged lepton sector. 
If there is quark--lepton--unification and it is the CKM matrix which 
corrects the tribimaximal mixing scheme, then almost 
maximal $CP$ violation and 
sizable deviation from zero $U_{e3}$ are implied. 
\end{abstract}

\newpage

\section{\label{sec:intro}Introduction}
Neutrino physics aims to fully determine all parameters of the neutrino 
mass matrix \cite{S}. Six out of the 9 physical low energy 
parameters are contained in the Pontecorvo--Maki--Nakagawa--Sakata (PMNS) 
neutrino mixing matrix \cite{PMNS}. It is given in general by 
$U = U_\ell^\dagger \, U_\nu$, where $ U_\nu$ diagonalizes the 
neutrino mass matrix and $U_\ell$ is associated to the diagonalization 
of the charged lepton mass matrix. 
The PMNS matrix can be parametrized as 
\bea \label{eq:Upara}
U = \left( \bad 
c_{12} c_{13} & s_{12} c_{13} & s_{13} e^{-i \delta}  \\[0.2cm] 
-s_{12} c_{23} - c_{12} s_{23} s_{13} e^{i \delta} 
& c_{12} c_{23} - s_{12} s_{23} s_{13} e^{i \delta} 
& s_{23} c_{13}  \\[0.2cm] 
s_{12} s_{23} - c_{12} c_{23} s_{13} e^{i \delta} & 
- c_{12} s_{23} - s_{12} c_{23} s_{13} e^{i \delta} 
& c_{23} c_{13}  \\ 
               \ea   \right) 
 {\rm diag}(1, e^{i \alpha}, e^{i (\beta + \delta)}) \, , 
\eea
where we have used the usual
notations $c_{ij} = \cos\theta_{ij}$, 
$s_{ij} = \sin\theta_{ij}$, 
$\delta$ is the Dirac $CP$ violation phase, 
$\alpha$ and $\beta$ are two possible Majorana 
$CP$ violation phases \cite{BHP80}. 
The values of the currently known mixing 
parameters are at $3\sigma$ \cite{valle,new_SC}: 
\begin{eqnarray} \label{eq:data1}
\sin^2 \theta_{12} & = & 0.24 \ldots 0.41 \; ,
\nonumber \\
|U_{e3}|^2 & \le & 0.044 \; ,
 \\
\sin^2 \theta_{23} & = & 0.34 \ldots 0.68\; .
\nonumber 
\end{eqnarray}
The probably most puzzling aspect of those numbers is that two of 
the mixing angles are compatible with two extreme values, 
namely zero for $\theta_{13}$ and 
$\pi/4$ for $\theta_{23}$. If confirmed by future data, this will be 
a strong hint towards the presence of special non--trivial symmetries in the 
lepton sector. 

\noindent 
Regarding the neutrino masses, we neither do know what the 
precise value of the lightest mass state is, 
nor do we know whether neutrinos are normally or inversely ordered. 
Instead, we have some information on the mass squared differences 
whose ratio is, again at $3\sigma$, given by \cite{valle}
\be \label{eq:data2} 
R \equiv  \frac{\Delta m^2_{\odot}}{\Delta m^2_{\rm A}} 
= \frac{m_2^2 - m_1^2}{\left| m_3^2 - m_1^2 \right|} 
= \frac{(7.1 \ldots 8.9) \cdot 10^{-5}}{(1.4 \ldots 3.3)\cdot 10^{-3}}
= 0.0215 \ldots 0.0636 \; .
\ee 
Many models have been constructed \cite{revs} in order to explain the mass and 
mixing schemes as implied by the data.
A particularly interesting mixing scenario, which is 
compatible with all current data, is 
the so--called tribimaximal scenario 
\cite{tri,devtri0,devtri11,devtri12}, defined by the following 
PMNS matrix: 
\be \label{eq:Utri} 
U = \left(
\bad 
\sqrt{\frac{2}{3}} & \sqrt{\frac{1}{3}} & 0 \\[0.2cm]
-\sqrt{\frac{1}{6}} & \sqrt{\frac{1}{3}} & \sqrt{\frac{1}{2}}  \\[0.2cm]
\sqrt{\frac{1}{6}} & -\sqrt{\frac{1}{3}} & \sqrt{\frac{1}{2}}  
\ea 
\right)~. 
\ee
It corresponds to 
$\sin^2 \theta_{12} = 1/3$, $U_{e3}=0$ and $\theta_{23}=\pi/4$. 
These numbers are referred to as ``tribimaximal values'' in this paper. 
Many authors have considered this scenario\footnote{Originally, 
a very similar, but with recent data incompatible form has been proposed 
already in \cite{tri1}.} from various points 
of view \cite{tri,devtri0,devtri11,devtri12}. We would like to 
note here that the most recent SNO 
salt phase data \cite{sno2} shifted the central value 
of $\sin^2 \theta_{12}$ to slightly 
higher values (the best--fit point went from 0.29 to 0.31 \cite{new_SC}). 
In addition, the allowed range of $\sin^2 \theta_{12}$ is now centered 
around the value 1/3 more than before 
(from $0.21 \ldots 0.38$ or $1/3\,^{+0.05}_{-0.12}$ 
to $0.24 \ldots 0.41$ or $1/3\,^{+0.08}_{-0.09}$). Hence, tribimaximal 
mixing is more than ever an interesting mixing scheme to investigate.\\    

\noindent 
In the future one expects some improvement on our knowledge of the 
neutrino mixing parameters (for summaries and more references, see 
for instance \cite{fut}). For our purposes, we mention that 
(at 3$\sigma$) 
next generation reactor and long--baseline experiments will probe 
$U_{e3}$ ($\sin^2 \theta_{23} - 1/2$) down to the 5 $\%$ (10 $\%$) 
level \cite{chef}. Further precision requires far--future projects 
such as neutrino factories or beta--beams. 
Low energy solar neutrino experiments \cite{pp} 
can probe $\sin^2 \theta_{12}$ with an uncertainty of $\sim 15 \%$, 
whereas more precision could stem from reactor experiments with a baseline 
corresponding to an oscillation minimum \cite{SPMIN}. \\

\noindent In this letter we wish to analyze the predictions of 
two different methods to perturb the tribimaximal mixing scenario. 
Instead of constructing models which realize such corrections, we rather 
give a summary of many possible scenarios. This will hopefully 
give a feeling for the expected values (in particular) 
of $\theta_{12, 13, 23}$ which can follow from 
models leading to (approximate) tribimaximal mixing. 
First, we assume in Section \ref{sec:mass} 
that the symmetry basis coincides with the basis in which the 
charged lepton mass matrix is real and diagonal. We then 
introduce at different positions in the mass matrix 
small breaking parameters and analyze 
the resulting predictions for the oscillation parameters. 

\noindent
The second approach lies in the impact of contributions from 
the charged lepton sector via $U_\ell$ to $U_\nu$. The latter matrix is 
assumed to correspond to tribimaximal mixing. This is discussed in 
Section \ref{sec:mix}. 
We present our conclusions and final remarks 
in Section \ref{sec:concl}.

\section{\label{sec:mass}Perturbing the Mass Matrix}
We shall work in this Section 
in a basis in which the charged leptons are real and diagonal, 
i.e., $U_\ell = \mathbbm{1}$. 
In general, a neutrino mass matrix leading to the predictions 
$\sin^2 \theta_{12} = 1/3$, $U_{e3}=0$ and $\theta_{23}=\pi/4$ 
has the form 
\be \label{eq:mnu}
m_\nu = 
\left(
\bad 
A & B & -B \\[0.2cm]
\cdot & \frac{1}{2} (A + B + D) & \frac{1}{2} (D - A - B)\\[0.2cm]
\cdot & \cdot & \frac{1}{2} (A + B + D)
\ea 
\right)~, 
\ee
where 
\be \label{eq:ABD}
A = \frac{1}{3} (2 \, m_1 + e^{i \alpha} \, m_2)~,~~
B = \frac{1}{3} (e^{i \alpha} \, m_2 - m_1)~,~~
D = e^{i \beta} \, m_3~, 
\ee
or $m_1 = A - B$ and $e^{i \alpha} \, m_2 = A + 2 \, B$. 
Interestingly, removing $B$ from the $\mu$--$\tau$ block of 
Eq.\ (\ref{eq:mnu}) will give  $\sin^2 \theta_{12} = 1/2$, i.e., exactly 
bimaximal mixing\footnote{In this case we would have 
$A = (m_1 + e^{i \alpha} \, m_2)/2$, $B = 
(e^{i \alpha} \, m_2 - m_1)/\sqrt{8}$ and $D$ as before.} 
\cite{bimax}. 
As well known, both the tribimaximal and the bimaximal 
mixing scheme are special cases of $\mu$--$\tau$ symmetric matrices, whose 
generic predictions are $U_{e3}=0$ and 
$\theta_{23}=\pi/4$ \cite{mutau,ma,Grimus:2004cc}.

\noindent Depending on the neutrino mass spectrum, the above neutrino mass 
matrix Eq.\ (\ref{eq:mnu}) simplifies further: 
\begin{itemize}
\item in case of an extreme normal hierarchy (NH) we have $m_1 \simeq 0$ 
and consequently 
$A \simeq B$ with $|A|, |B| \ll |D|$;  
\item in case of an extreme inverted hierarchy (IH) 
we have $m_3 = D \simeq 0 $. If in addition $m_2$ and $m_1$ have opposite 
(identical) $CP$ parities, one has $A \simeq \sqrt{\dma}/3$ and 
$B \simeq -2\sqrt{\dma}/3$ ($A \simeq \sqrt{\dma}$ and $B \simeq 0$); 
\item in case of a quasi--degenerate spectrum we can neglect the mass 
splittings and have for equal $CP$ parities of $m_1$ and $m_2$ that 
$B \simeq 0$ and $|A| \simeq |D|$. If in addition $m_3$ has the same parity as 
$m_1$ and $m_2$, then the mass matrix is proportional to the unit matrix: 
$m_\nu = A \, \mathbbm{1} \simeq m_0 \, \mathbbm{1} $. 
If  $m_1$ and $m_2$ have opposite parities, then $A \simeq m_0/3 \simeq -B/2$ 
and $|A| \simeq |D|/3$. If sgn($m_1$)= sgn($m_2$)=$-$sgn($m_3$), then 
$A \simeq -D$, $B=0$ and all elements of the 
mass matrix are zero except for the $ee$ entry $A$, and the 
$\mu\tau$ entry $-A$. This matrix conserves the flavor charge 
$L_\mu - L_\tau$ \cite{lmlt} and has for instance been obtained also in 
Refs.\ \cite{A4}. 
\end{itemize}
In this letter we shall focus on the normal and inverted hierarchical case, 
since radiative corrections, in particular for $\theta_{12}$, 
are in general very strong for quasi--degenerate neutrinos.\\

\noindent 
In the following Subsections we shall analyze perturbations of 
the tribimaximal mass matrix. With perturbations we 
mean that an element of the matrix 
Eq.\ (\ref{eq:mnu}) is multiplied with $(1 + \epsilon)$, 
where $\epsilon \ll 1$. These corrections might stem for instance from 
higher dimensional Higgs representations of the underlying theory which 
generates tribimaximal mixing or by the fact that the theory does 
not exactly reproduce the matrix in Eq.\ (\ref{eq:mnu}). 
There are two cases which result in deviations of all three 
mixing parameters, namely perturbations in the 
$e\tau$ element or in the $\mu\mu$ element. They are equivalent 
to perturbations in the $e\mu$ element or in the $\tau\tau$ element 
when the sign of $\epsilon$ is changed. 
Alternatively, one can perturb the $ee$ or $\mu\tau$ element of $m_\nu$ 
and one will find that only the relation 
$\sin^2 \theta_{12} = 1/3$ is altered, whereas 
$U_{e3}=0$ and $\theta_{23} = \pi/4$ remain. 
This is because the $\mu$--$\tau$ symmetry is unchanged when the 
$ee$ or $\mu\tau$ element are altered.  
In Section \ref{sec:NH} and \ref{sec:IH} we analyze these possibilities 
in case of NH and IH, respectively. 
The case of applying radiative corrections to the mass matrix 
is treated separately in Section \ref{sec:RGE}.

\subsection{\label{sec:NH}Normal Hierarchy}
Let us consider first the case of 
normal hierarchy, i.e., $|D| \gg |A|,  |B|$. 
It is known from the study of $\mu$--$\tau$ symmetric matrices that 
if the symmetry is broken in the electron sector, $|U_{e3}|$ deviates 
more from zero than $\theta_{23}$ does from $\pi/4$ \cite{rabimutau}. 
Indeed, the matrix which is obtained by perturbing the $e\tau$ element, 
\be \label{eq:NHmnu_e}
m_\nu' = 
\left(
\bad 
A & B & -B(1 + \epsilon) \\[0.2cm]
\cdot & \frac{1}{2} (A + B + D) & \frac{1}{2} (D - A - B)\\[0.2cm]
\cdot & \cdot & \frac{1}{2} (A + B + D)
\ea 
\right)~, 
\ee
predicts that 
$\theta_{23}$ remains rather close to maximal, 
$\theta_{23} = \pi/4 + {\cal O}(\epsilon^2)$, and 
\be \label{eq:NHresmnu_e}
\sin^2 \theta_{12} \simeq \frac{1}{3} 
\left(1 + \frac{2}{9} \, \epsilon \right) 
~\mbox{ and } ~
|U_{e3}| \simeq \frac{\epsilon}{\sqrt{2}} \frac{B}{D} \simeq 
\frac{m_2 - m_1}{3 \sqrt{2} \, m_3} \, \epsilon \simeq 
\frac{\sqrt{R}}{3 \, \sqrt{2} } \, \epsilon ~.
\ee 
Note that the parameters $A,B,D$ drop out of the expression 
for $\sin^2 \theta_{12}$, leading to a sharp prediction for 
solar neutrino mixing as a function of $\epsilon$.  
The parameter $|U_{e3}|$ is suppressed by the small breaking 
parameter $\epsilon$ and by the ratio $(m_2 - m_1)/m_3$, which is roughly the  
square root of the solar and atmospheric $\Delta m^2$. 
In total, one therefore expects that $|U_{e3}| \ls R$. 
We can write down the following correlations between the 
observables: 
\be
\sin^2 \theta_{12} \simeq \frac{1}{3} \left( 1 + \frac{2 \sqrt{2}}{3} \, 
\frac{|U_{e3}|}{\sqrt{R}} \right)~\mbox{ and }~ 
\meff \simeq \frac{\sqrt{\dms}}{3} \simeq 
\frac{2 \sqrt{2}}{27} \frac{|U_{e3}| \, 
\sqrt{\dma}}{\sin^2 \theta_{12} - \frac{1}{3} }~.
\ee
The last equation holds in the limit $A \simeq B$, which corresponds 
to a very strong hierarchy of the neutrino masses. 
Of course, the effective mass in case of NH is very small.\\

\noindent Now we perturb the tribimaximal mixing 
scenario in the $\mu\mu$ element, such that 
\be \label{eq:NHmnu_mt}
m_\nu' = 
\left(
\bad 
A & B & -B \\[0.2cm]
\cdot & \frac{1}{2} (A + B + D) (1 + \epsilon)
& \frac{1}{2} (D - A - B)\\[0.2cm]
\cdot & \cdot & \frac{1}{2} (A + B + D)
\ea 
\right)~. 
\ee
Then it holds for $A \simeq B$ that 
\be \label{eq:NHresmnu_mt}
\sin^2 \theta_{23} \simeq \frac{1}{2} \left(1 + \frac{\epsilon}{2} \right)~,~
|U_{e3}| \simeq \frac{\epsilon}{2 \sqrt{2}} \, \frac{B}{D}~. 
\ee 
In contrast to the case of breaking in the $e$ sector, we have  
in the expressions for the mass eigenstates now terms of 
order $B$ and $D \epsilon$, which might be all of the same order. 
This prohibits to directly connect $B/D$ with the ratio of the 
mass--squared differences. Nevertheless, $B/D$ is of the order of 
$\sqrt{R}$, so that a similar magnitude for $|U_{e3}|$ as in case 
of a perturbation in the $e\tau$ element is expected. 
For the solar neutrino mixing angle we find 
\be \label{eq:NHresmnu_mt1}
\tan 2 \theta_{12} \simeq \frac{2\sqrt{2}}
{1 + D/B \, \epsilon/4} ~, 
\ee
where in contrast to the previous case the parameters $B$ and $D$, which are 
connected to the masses, appear. Thus, as a function of 
$\epsilon$ there will now 
be a sizable spread in the allowed values of $\sin^2 \theta_{12}$. 
We can rewrite the last formulae as 
\be
\tan 2 \theta_{12} \simeq \frac{2 \sqrt{2}}
{1 + (\sin^2 \theta_{23} - 1/2)^2 /(8 \sqrt{2} \, |U_{e3}|) 
}
\simeq \frac{2\sqrt{2}}
{1 +(\sin^2 \theta_{23} - 1/2) \,  \sqrt{\dma}/\meff  }~, 
\ee
where the last estimate is valid for $A \simeq B$.  
From the 
approximative expressions for the mixing parameters, we expect that 
both the deviations of $\sin^2 \theta_{12}$ from $1/3$ and of 
$\sin^2 \theta_{23}$ from $1/2$ are larger when 
breaking occurs in the $\mu\mu$ element. 

\noindent 
We relax now the assumption of an extreme hierarchy $A \simeq B \ll D$ 
and consider the case that $|A|$ and $|B|$ are of the same order and 
one order of magnitude smaller than $|D|$. Allowing further that those 
quantities are complex, we turn to a numerical analysis. 
Fig.\ \ref{fig:NH} shows the results, which confirm our 
analytical estimates from above.  
In order to generate this plot we required the oscillation parameters to lie 
in their allowed 3$\sigma$ ranges given in Eqs.\ 
(\ref{eq:data1}, \ref{eq:data2}) 
and used the mass matrices from 
Eqs.\ (\ref{eq:NHmnu_e}, \ref{eq:NHmnu_mt}) with a fixed value of 
$\epsilon=0.1$. 
As can be seen from the approximate expressions in 
Eqs.\ (\ref{eq:NHresmnu_e}, \ref{eq:NHresmnu_mt}, \ref{eq:NHresmnu_mt1}), 
the direction in which the deviations of 
$\sin^2 \theta_{12}$ and $\sin^2 \theta_{23}$ 
from their tribimaximal values go depend on the sign of $\epsilon$. 
This change of sign is equivalent to a relocation of the perturbation. 
For instance, for positive $\epsilon$ and a perturbation in the 
$e\tau$ element, we have $\sin^2 \theta_{12} > 1/3$. 
If $\epsilon$ is negative or we break the symmetry in the $e\mu$ element 
with positive $\epsilon$,  we would have $\sin^2 \theta_{12} < 1/3$.\\  

\noindent 
Other breaking scenarios may be possible, for instance, one might 
perturb only the $ee$ element of Eq.\ (\ref{eq:mnu}). 
Then $U_{e3}$ and $\theta_{23}$ will keep their values 
0 and $\pi/4$, respectively, and only the solar neutrino mixing angle will be 
modified, according to 
\be \label{eq:ee}
\sin^2 \theta_{12} \simeq \frac{1}{3} \left( 1  
+ \frac{4}{9} \frac{A}{B} \, \epsilon \right)~, 
\ee
where $A/B$ is approximately equal to one for NH. 
The deviation is therefore small.  
Perturbing the $\mu\tau$ entry of $m_\nu$ also leaves 
$U_{e3}$ and $\theta_{23}$ unchanged and leads to 
\be \label{eq:mt}
\sin^2 \theta_{12} \simeq \frac{1}{3} \left( 1  
- \frac{2}{9} \frac{A + B - D}{B} \, \epsilon \right)~.
\ee 
The deviation from 
1/3 is rather sizable, since $(A + B - D)/B \sim 1/\sqrt{R}$. 
Hence, $\sin^2 \theta_{12}$ lies typically close to its 
maximal allowed value.

\subsection{\label{sec:IH}Inverted Hierarchy}
Now we turn to the inverted hierarchy, where the requirement of 
stability under renormalization motivates us to study the case 
$A \simeq -B/2$. This can be understood from Eq.\ (\ref{eq:ABD}) which 
states that for $A \simeq -B/2$ the two leading mass states possess 
opposite $CP$ parity. This stabilizes the oscillation parameters 
connected to solar neutrinos \cite{RGE}. 
For effects solely attributed to radiative corrections, see Section 
\ref{sec:RGE}. 
First we consider breaking in the $e\tau$ entry of Eq.\ (\ref{eq:mnu}), i.e., 
the matrix from Eq.\ (\ref{eq:NHmnu_e}). 
Setting $D=0$ we have 
\be \label{eq:IHresmnu_e}
\sin^2 \theta_{12} \simeq \frac{1}{3} \left(1 + \frac{2}{9} \, \epsilon
\right)~,~|U_{e3}| \simeq \frac{A + B}{2\sqrt{2} \, B} \, \epsilon
\simeq \frac{1}{4 \sqrt{2}} \, \epsilon \simeq \sqrt{\frac{3}{14}} 
\, \sqrt{R}~, 
\ee 
where the last estimates for $|U_{e3}|$ are valid for $A \simeq -B/2$. 
In this case we also have $\sin^2 \theta_{12} \simeq 1/3~(1 + 
8\sqrt{2} \, \epsilon /9)$. 
In contrast to the case of normal hierarchy given above, atmospheric neutrino 
mixing can deviate significantly from maximal, namely 
\be 
\sin^2 \theta_{23} \simeq \frac{1}{2} (1 + \epsilon) \simeq 
\frac{1}{2} \left(1 + 4 \sqrt{2} \, |U_{e3}| \right)~.
\ee 
The parameters $A,B,D$ drop out of the expressions for $\sin^2 \theta_{12} $ 
and $\sin^2 \theta_{23}$, leading to a sharp 
prediction for this observable.\\

\noindent 
The next case is to break the symmetry in the $\mu\mu$ element, i.e., 
we consider Eq.\ (\ref{eq:NHmnu_mt}) with $D \ll A \simeq -B/2$. 
As a result, 
\be
|U_{e3}| \simeq \frac{A + B}{4 \sqrt{2} \, B} \, \epsilon 
\simeq \frac{1}{8\sqrt{2}} \, \epsilon~,~
\sin^2 \theta_{23} \simeq \frac{1}{2} \left( 
1 - \frac{A (A + B)}{2\sqrt{2} \, B} \, \epsilon
\right) \simeq \frac{1}{2}
\left( 
1 - \frac{1}{8\sqrt{2}} \, \epsilon
\right)~,
\ee
which shows that $U_{e3}$ and $\sin^2 \theta_{23}$ witness 
only small corrections. 
What regards the solar neutrino mixing parameter, 
\be 
\sin^2 \theta_{12} \simeq \frac{1}{3} \left( 
1 - \frac{A + B}{9 \, B}  \, \epsilon  
\right) \simeq \frac{1}{3} \left( 1 - \frac{1}{18} \, \epsilon \right)~.
\ee
As already noted in \cite{rabimutau}, the correlation between 
$|U_{e3}|$ and $R$ is not particularly strong or even absent in case 
of IH. The estimates for the solar neutrino parameters are not very 
reliable in this case. 
Again, we present a numerical analysis of the general consequences. 
We assume that $|D|$ is one order of magnitude smaller than $|A|, |B|$, 
which in turn fulfill $|A| \simeq -|B|/2$. 
Diagonalizing with such values 
the matrices from (\ref{eq:NHmnu_e}, \ref{eq:NHmnu_mt}) gives 
Fig.\ \ref{fig:IH} (again for $\epsilon = 0.1$), 
which confirms our analytical estimates. 
We also display the ratio $\dma/\meff$ in this Figure. Note that with 
opposite parities of the two mass states in case of IH it holds that 
$\meff \simeq \sqrt{\dma} \, \cos 2 \theta_{12}$, where 
for exact tribimaximal mixing 
$\cos 2 \theta_{12} = \sin^2 \theta_{12} = 1/3$.  
Again, changing the sign of 
$\epsilon$ (or perturbing the matrix in the $e\mu$ ($\tau\tau$) 
instead of the $e\tau$ ($\mu\mu$) element) 
leads approximately to a reflection of $\sin^2 \theta_{12}$ and 
$\sin^2 \theta_{23}=1/2$ around their tribimaximal values.\\

\noindent 
One can again perturb the $ee$ or the $\mu\tau$ element, which does not change 
$U_{e3}=0$ or $\sin^2 \theta_{23}=1/2$ but alters $\theta_{12}$ 
according to Eqs.\ (\ref{eq:ee}, \ref{eq:mt}). In our case of 
opposite $CP$ parities we have for a perturbation in the $ee$ element that  
$\sin^2 \theta_{12} \simeq \frac{1}{3} 
+ \frac{4}{27} \frac{A}{B} \, \epsilon $ with $A/B \simeq -1/2$. 
Perturbing the $\mu\tau$ element also gives negligible corrections, namely 
$\sin^2 \theta_{12} \simeq \frac{1}{3} - \epsilon/27$. 

\subsection{\label{sec:RGE}Radiative Corrections}
Radiative corrections \cite{RGE} 
can also play a role to generate deviations 
from tribimaximal mixing. Running from a high scale $M_X \simeq 10^{16}$ GeV 
at which the tribimaximal mass matrix is assumed to be generated to the 
low scale has an effect on the mass matrix Eq.\ (\ref{eq:mnu}) of 
\be \label{eq:mnuRG}
m_\nu' = 
\left(
\bad 
A & B & -B (1 + \epsilon)  \\[0.2cm]
\cdot & \frac{1}{2} \left( A + B + D \right) 
& \frac{1}{2} (D - A - B)(1 + \epsilon) \\[0.2cm]
\cdot & \cdot & \frac{1}{2} (A + B + D)(1 + 2 \epsilon) 
\ea 
\right),\mbox{where } 
\epsilon = c \, \frac{m_\tau^2}{16 \pi^2 \, v^2} \ln \frac{M_X}{m_Z} ~.
\ee
The parameter $c$ is given by 3/2 in the SM and by $-(1 + \tan^2 \beta)$ 
in the MSSM. We find that in the SM there are no sizable effects on the 
observables. In case of the MSSM, however, interesting effects occur 
for NH and $\tan \beta \gs 30$. 
We calculate in case of a normal hierarchy the observables to be 
\be
|U_{e3}| \simeq \frac{\sqrt{2} B }{D} \epsilon ~,~~
\sin^2 \theta_{23} \simeq \frac{1}{2} (1 - \epsilon)~\mbox{ and }~
\sin^2 \theta_{12} \simeq \frac{1}{3} \left( 1 - \frac{2}{9} (1 + 2 \, A/B)
\, \epsilon \right)~.
\ee
With negative $\epsilon$, as in the case of the relevant corrections 
in the MSSM, we have $\sin^2 \theta_{23}$ above $1/2$ and 
$\sin^2 \theta_{23}$ above $1/3$.\\ 

\noindent 
Considering IH we have two cases, corresponding to 
$|A| \simeq |B|/2 \gg |D|$ and $|A| \gg |B|, |D|$. 
The first possibility implies opposite $CP$ parities of the 
leading mass states and the second one 
identical parities, which is known to be unstable under 
radiative corrections \cite{RGE}. It turns out that 
in this case there is indeed an  
upper limit on $\tan \beta$, given by roughly $\tan \beta \gs 15$, i.e., 
larger values are not compatible with the data.   
Both cases imply that $U_{e3}$ is small, namely of order $\epsilon^2$ 
or $\epsilon \, D/A$. 
For $|A| \simeq |B|/2 \gg |D|$ we find that  
\be
\sin^2 \theta_{12} \simeq \frac{1}{3} \left( 1 - \frac{2}{9 \, B} 
(2 A + B)\, \epsilon \right) ~\mbox{ and }~
\sin^2 \theta_{23} \simeq \frac{1}{2} \left(1 + \epsilon \right)~. 
\ee
Therefore, with negative $\epsilon$ we have $\sin^2 \theta_{12} > 1/3$ 
and  $\sin^2 \theta_{12} < 1/2$. The same is implied for the rather 
unstable case of $|A| \gg |B|, |D|$, for which we find 
\be
\tan 2\theta_{12} \simeq 
\frac{2\sqrt{2}~(1 + \epsilon)}{1 + (1 + A/B \, \epsilon)} ~\mbox{ and }~
\sin^2 \theta_{23} \simeq \frac{1}{2} ( 1 + \epsilon)~. 
\ee
As expected, $\theta_{12}$ receives sizable corrections, since 
$A/B \sim 1/\sqrt{R}$. 
Interestingly, $\sin^2 \theta_{12}$ is always larger than 1/3 when radiative 
corrections are applied. 
We plot in Fig.\ \ref{fig:RG} some of the resulting scatter plots. 
The parameter $U_{e3}$ is below 0.01 in all three cases. \\

\section{\label{sec:mix}Perturbing the Mixing Matrix}
The second possibility to deviate a zeroth order mixing scheme is by 
taking correlations from the charged lepton sector into account. 
This approach has been followed mainly to analyze deviations from 
bimaximal neutrino mixing \cite{devbimax,QLC,FPR}, but also has 
been studied in case of 
deviations from tribimaximal mixing \cite{devtri11,devtri12}, 
where similar studies can be found.  
We note here that the analysis in this Section is independent on the 
mass hierarchy of the neutrinos. 
In general, the PMNS matrix is given by
\be
U = U^\dagger_{\ell} \, U_\nu ~, 
\ee
where $U_{\ell}$ is the $3\times3$ unitary matrix associated with the 
diagonalization of the charged lepton mass matrix and 
$U_\nu$ diagonalizes the neutrino Majorana mass term. 
As has been shown in Ref.\ \cite{FPR}, one can in general express the 
PMNS matrix as 
\be\label{eq:parapmns}
U =  
\tilde{U}_{\ell}^\dagger \, P_\nu \, \tilde{U}_\nu \, Q_\nu ~.
\ee
It consists of two diagonal phase matrices 
$P_\nu$ = diag($1,e^{i \phi}, e^{i \omega}$) and 
$Q_\nu$ = diag($1,e^{i \rho}, e^{i \sigma}$), as well as two 
``CKM--like'' matrices $\tilde{U}_\ell$ and $\tilde{U}_\nu$ which contain 
three mixing angles and one phase each, and are parametrized in analogy to 
Eq.\ (\ref{eq:Upara}). All matrices except for $\tilde{U}_\ell$ stem from the 
neutrino sector and out of the six phases present in 
Eq.\ (\ref{eq:parapmns}) five stem from the neutrino 
sector. We denote the phase in $\tilde{U}_{\ell}$ with $\psi$. 
The six phases will in general 
contribute in a complicated manner to the three observable ones. 
 In case of one angle in $U_\nu$ being zero, the 
phase in $\tilde{U}_\nu$ is unphysical. 
Note that the 2 phases in $Q_\nu$ do not appear in 
observables describing neutrino oscillations \cite{FPR}. 
We assume in the following that $U_\nu$ corresponds to 
tribimaximal mixing, i.e., to Eq.\ (\ref{eq:Utri}).\\

\noindent Let us define $\lambda_{ij}$ to be the 
sine of the mixing angles of $U_\ell$, 
i.e., $\lambda_{ij} = \sin \theta_{ij}^\ell$.  We can assume that 
the $\lambda_{ij}$ are small and express the neutrino observables as 
functions of the $\lambda_{ij}$. We find: 
\bea \label{eq:ser_obs_1}
\sin^2 \theta_{12} \simeq \frac{1}{3} - \frac{2}{3} \, c_\phi \, \lambda_{12} 
+ \frac{2}{3} \, c_{\omega - \psi} \, \lambda_{13} ~,\\[0.3cm]
|U_{e3}| \simeq \frac{1}{\sqrt{2}} \,  
\left| \lambda_{12} + e^{i(\omega - \psi - \phi)} \, \lambda_{13} \right|
~,\\[0.3cm]
\sin^2 \theta_{23} \simeq \frac{1}{2} - c_{\omega - \phi} \, \lambda_{23} ~, 
\eea 
where we neglected terms of order $\lambda_{ij}^2$. 
We introduced the notations 
$c_\phi=\cos\phi$, $c_{\omega - \psi} = \cos(\omega - \psi)$ and so on.
The parameter 
$\lambda_{23}$ does in first order not appear in $\sin^2 \theta_{12}$ and 
$|U_{e3}|$, just as $\lambda_{12,13}$ does not in $\sin^2 \theta_{23}$. 
The observed smallness of the deviation of $\sin^2 \theta_{12}$ from 1/3 
indicates that the $\lambda_{ij}$ are very small. 

\noindent 
There are two kinds of invariants parameterizing $CP$ violating effects. 
First, we have the Jarlskog invariant \cite{CJ85} describing all $CP$ breaking 
observables in neutrino oscillations \cite{PKSP3nu88}. 
It is given by 
\be \label{eq:ser_jcp_1}
J_{CP} = {\rm Im} 
\left\{ 
U_{e1} \, U_{\mu 2} \, U_{e 2}^\ast \, U_{\mu 1}^\ast 
\right\} \simeq 
\frac{1}{6} s_\phi \, \lambda_{12} + \frac{1}{6} s_{\omega - \phi} 
\, \lambda_{13}~. 
\ee 
Again, the notation is $s_\phi = \sin \phi$, $s_{\omega - \phi} = 
\sin (\omega - \phi)$ and so on.  
In the parametrization of Eq.\ (\ref{eq:Upara}) one has 
$J_{CP} = \frac{1}{8} \sin 2 \theta_{12} \, \sin 2 \theta_{23} \, 
\sin 2 \theta_{13} \, \cos \theta_{13} \, \sin \delta $, which for 
$\sin^2 \theta_{12} = 1/3$, maximal $\theta_{23}$ and small $U_{e3}$ 
simplifies to $J_{CP} \simeq \frac{1}{3\sqrt{2}} \, |U_{e3}| \, \sin \delta $. 
The phase $\phi$ is in leading order the phase governing 
low energy $CP$ violation in neutrino oscillation experiments. 
Note though that this does not necessarily 
imply\footnote{We thank Steve King and 
Stefan Antusch for discussions on this point.} that $\phi$  
is identical to the Dirac phase $\delta$ 
in the parametrization of the PMNS matrix in Eq.\ (\ref{eq:Upara}).

\noindent 
Then there are two invariants $S_1$ and $S_2$ \cite{JMaj87}, 
only valid for $U_{e3}\neq 0$, which are 
related to the two Majorana phases $\alpha$ and $\beta$. 
They are 
\bea \label{eq:ser_S_1}
S_1 = {\rm Im}\left\{ U_{e1} \, U_{e3}^\ast \right\} \simeq 
\frac{1}{\sqrt{3}} s_{\phi + \sigma} \, \lambda_{12} + 
\frac{1}{\sqrt{3}} s_{\omega - \psi + \sigma} \, \lambda_{13} ~,\\[0.3cm]
S_2 = {\rm Im}\left\{ U_{e2} \, U_{e3}^\ast \right\} 
\simeq 
\frac{1}{\sqrt{6}} s_{\phi - \rho + \sigma} \, \lambda_{12} + 
\frac{1}{\sqrt{6}} s_{\omega - \psi - \rho + \sigma} \, \lambda_{13}~.
\eea
It is seen that the two phases $\rho$ and $\sigma$ from $Q_\nu$ 
appear only in $S_1$ and $S_2$, which are associated with the Majorana 
phases.\\

\noindent Let us assume now a ``CKM--like'' hierarchy of the 
mixing angles in $U_\ell$, i.e., 
$\lambda_{12}=\lambda$, 
$\lambda_{23}=A \, \lambda^2$ and $\lambda_{13}=B \, \lambda^3$ 
with $A$, $B$ real and of order one. 
In this case the $CP$ conserving observables are given by: 
\bea \label{eq:ser_obs_2}
\sin^2 \theta_{12} \simeq \frac{1}{3} - \frac{2}{3} \, c_\phi \, \lambda  
+ \frac{1}{6} \left( c_{2 \phi} + s_{2 \phi} \right)  \lambda^2 ~,\\[0.3cm]
|U_{e3}| \simeq \frac{1}{\sqrt{2}} \,  \lambda ~,\\[0.3cm]
\sin^2 \theta_{23} \simeq \frac{1}{2} + \frac{1}{4} 
\left( c_{2 \phi} + s_{2 \phi} - 2 - 4A \, c_{\omega - \phi} 
\right)\, \lambda^2 ~,
\eea 
plus terms of ${\cal O}(\lambda^3)$. 
We see that unless there is sizable cancellation due to the 
phase $\phi$ we must have small $\lambda$. To first order we have 
from Eq.\ (\ref{eq:data1}) that $\frac{2}{3} \, c_\phi \, \lambda $ 
has to lie between $-0.08$ and $0.09$. 
For the $CP$ violating observables we have 
\bea \label{eq:ser_CP_2}
J_{CP} \simeq \frac{1}{6} s_\phi \, \lambda ~,\\[0.3cm]
S_1 \simeq \frac{1}{\sqrt{3}} s_{\phi + \sigma} \, \lambda + 
\frac{1}{2\sqrt{3}} s_\sigma \, \lambda^2 ~,\\[0.3cm]
S_2 \simeq \frac{1}{\sqrt{6}} s_{\phi - \rho + \sigma} \, \lambda + 
\frac{1}{\sqrt{6}} s_{\rho - \sigma} \, \lambda^2~, 
\eea
where again terms of order $\lambda^3$ are neglected\footnote{
For the special case of $\rho=\sigma=0$ we would have 
(to order $\lambda^3$) that 
$J_{CP}\simeq \frac{S_1}{2\sqrt{3}} \simeq \frac{S_2}{\sqrt{6}}$, 
and the magnitudes of the $CP$ violating effects in neutrino 
oscillations and the $CP$ violation associated with their 
Majorana nature are directly proportional.}.  
We see that the phase governing $CP$ violation in neutrino oscillations 
appears also in $\sin^2 \theta_{12}$. This phase has its origin in the 
neutrino sector. Note that due to the hierarchical structure of $U_{\ell}$ 
the phase $\psi$ does appear only at third order of $\lambda$ 
in the observables. 

\noindent 
We can obtain an interesting 
correlation from the expressions for the mixing parameters in 
Eq.\ (\ref{eq:ser_obs_2}):
\be \label{eq:cor}
\sin^2 \theta_{12} \simeq \frac{1}{3} - \frac{2\sqrt{2}}{3} \, c_\phi \, 
|U_{e3}|~.
\ee
A similar relation has been obtained in \cite{devtri12}.
Moreover, the deviation of $\sin^2 \theta_{23}$ from 1/2 is of order 
$|U_{e3}|^2$, which in turn is of order $(1/3 - \sin^2 \theta_{12})^2$. 
If one of the 
parameters $|U_{e3}|$, $\sin^2 \theta_{12}$ or $\sin^2 \theta_{23}$ is 
close to its tribimaximal value 0, 1/3 or 1/2, 
then automatically the others 
are as well. 

\noindent 
Consider now the case of sizable $\lambda$: 
we can see from Eqs.\ (\ref{eq:ser_obs_2}) and 
(\ref{eq:ser_CP_2}) that for sizable $\lambda$ the parameter $|U_{e3}|$ 
is also sizable. 
Since experimental data shows that 
$\sin^2 \theta_{12}$ is close to 1/3, the phase 
$\phi$ has to be around $\pi/2$ or $3\pi/2$ so that the first term in 
the expansion for $\sin^2 \theta_{12}$ vanishes. 
For such values of the phase, however, $J_{CP}$ is large. 
Hence, large $|U_{e3}|$ requires large $CP$ violation. 
This correlation between $|U_{e3}|$ and large $CP$ violation 
would be the case also if we identify $U_\ell$ with the 
CKM--matrix, i.e., $\lambda = \sin \theta_C = 0.22$, where $\theta_C$ is the 
Cabibbo angle. 
This scenario which has recently gathered 
some attention in case of bimaximal $U_\nu$ 
and received the name ``Quark--Lepton--Complementarity'' 
\cite{QLC}. In our case, if indeed $\lambda = \sin \theta_C$, 
then the predictions are (with $\phi \simeq \pm \pi/2$) that 
the deviation of $\sin^2 \theta_{12} $ from 1/3 is very small and 
that $\sin^2 \theta_{23}$ can differ only up to a few $\%$ from 1/2: 
\bea
U_{e3} \simeq 0.22/\sqrt{2} \simeq 0.16~,~
\sin^2 \theta_{12} \simeq \frac{1}{3} (1 - |U_{e3}|^2) \simeq 0.325 \\[0.3cm]
\mbox{ and } 
\sin^2 \theta_{23} \simeq \frac{1}{2} (1 \mp \frac{3}{2} 
A \, s_\omega \, |U_{e3}|^2 ) \simeq 0.49 \ldots 0.52~, 
\eea 
where $A \simeq 0.8$ was used. The phase $\omega$ in the expression for 
$\sin^2 \theta_{23}$, which determines the size of the deviation from 
maximal mixing, is unobservable. 

\noindent
In Fig.\ \ref{fig:mix} we see the result of an analysis of a  
scenario in which $U_\ell$ is ``CKM--like''. 
We vary $\lambda = \sin \theta_{12}^\ell$ between zero and 
0.25 and choose $\lambda_{23}=A \, \lambda^2$ and 
$\lambda_{13}=B \, \lambda^3$ with $A$ and $B$ between $1/\sqrt{3}$ 
and $\sqrt{3}$. Our analytical estimates from above are nicely confirmed.

\section{\label{sec:concl}Conclusions and final Remarks}
\begin{figure}[t]\vspace{-5cm}
\hspace{-3cm}
\epsfig{file=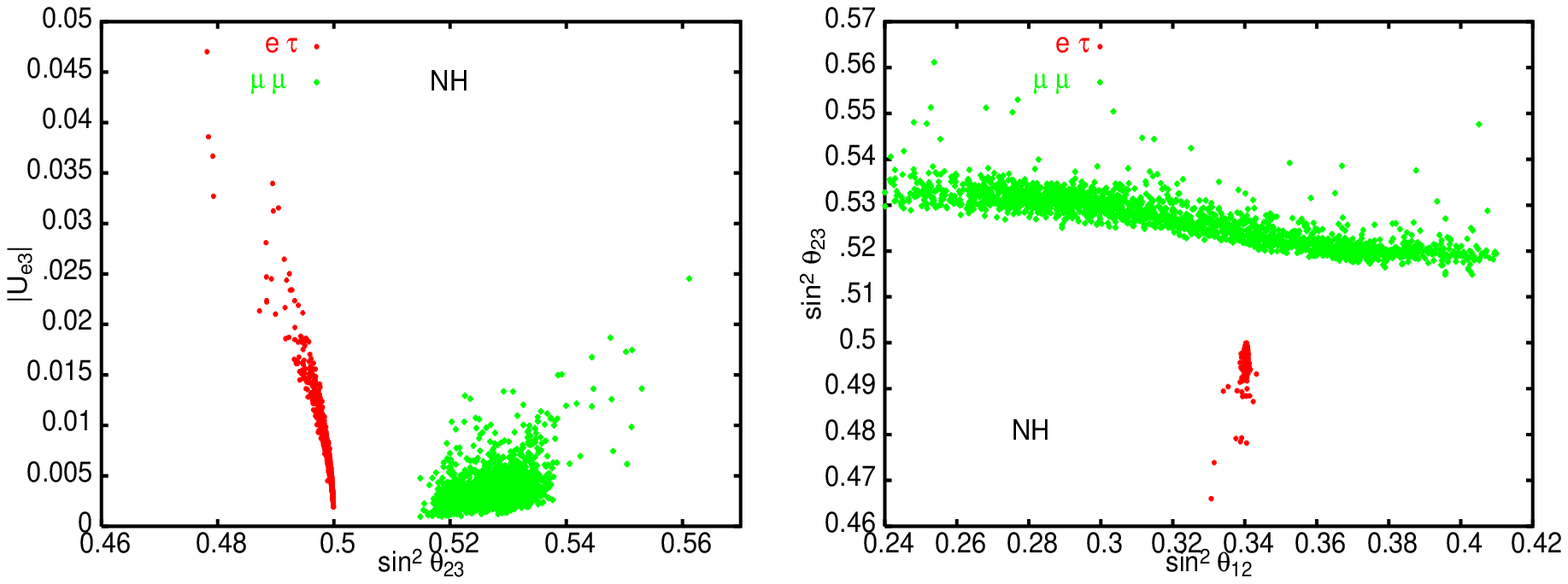,width=21cm,height=29cm}
\vspace{-17.4cm}
\caption{\label{fig:NH}Scatter plots of some of the correlations 
of the observables for the normal hierarchy 
resulting from Eqs.\ (\ref{eq:NHmnu_e}) and 
(\ref{eq:NHmnu_mt}). The parameters $A,B,D$ are complex 
and $|A|, |B|$ are smaller than $D$ by one order of magnitude. 
The dark (red) points are for a perturbation in the 
$e\tau$ element, the bright (green) points for a perturbation in the 
$\mu\mu$ element.
}\vspace{-2cm}
\hspace{-6cm}\vspace{-1.4cm}
\epsfig{file=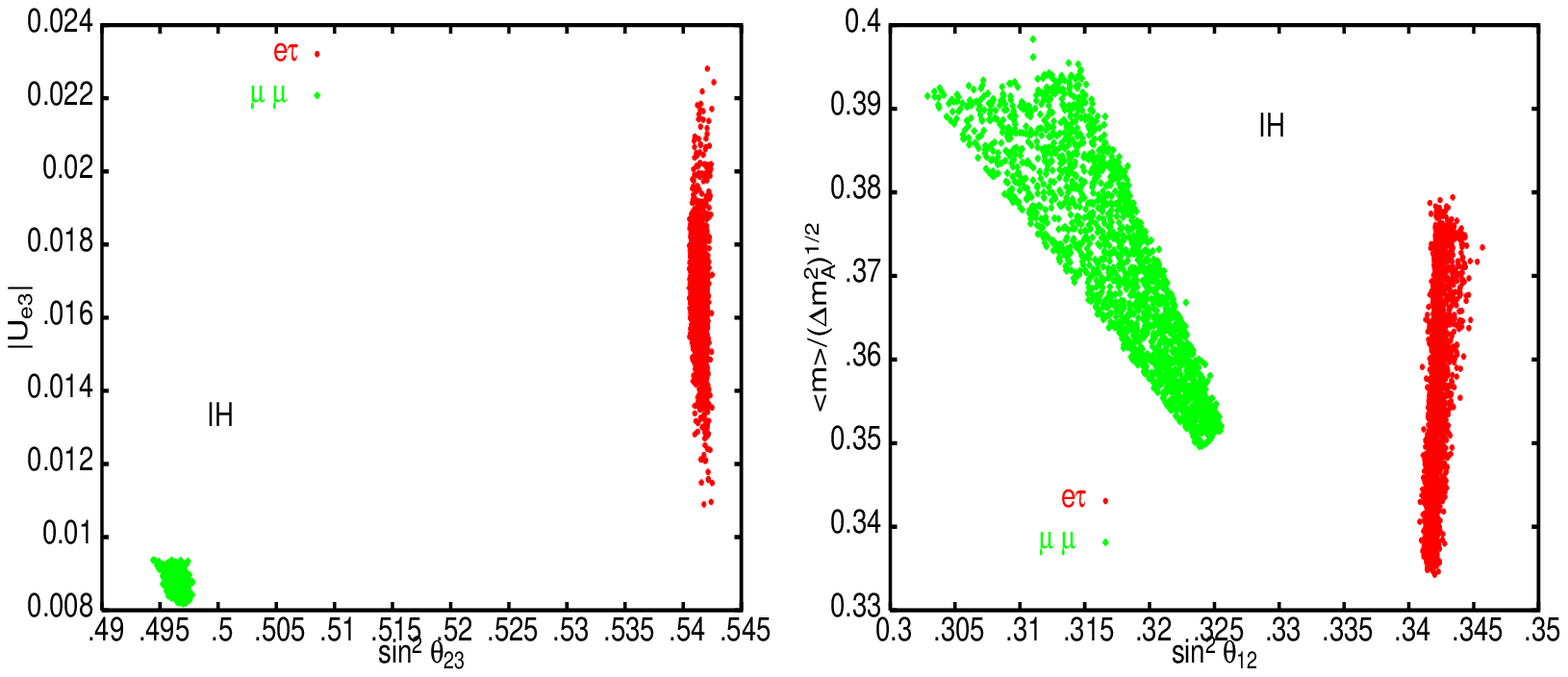,width=24cm,height=29cm}
\vspace{-18cm}
\caption{\label{fig:IH}Same as above for the inverted hierarchy. We have 
now $|D|$ smaller than $|A|, |B|$ by one order of magnitude and 
$|A| \simeq |B/2|$.}
\end{figure}

\begin{figure}[t]
\begin{center}
\epsfig{file=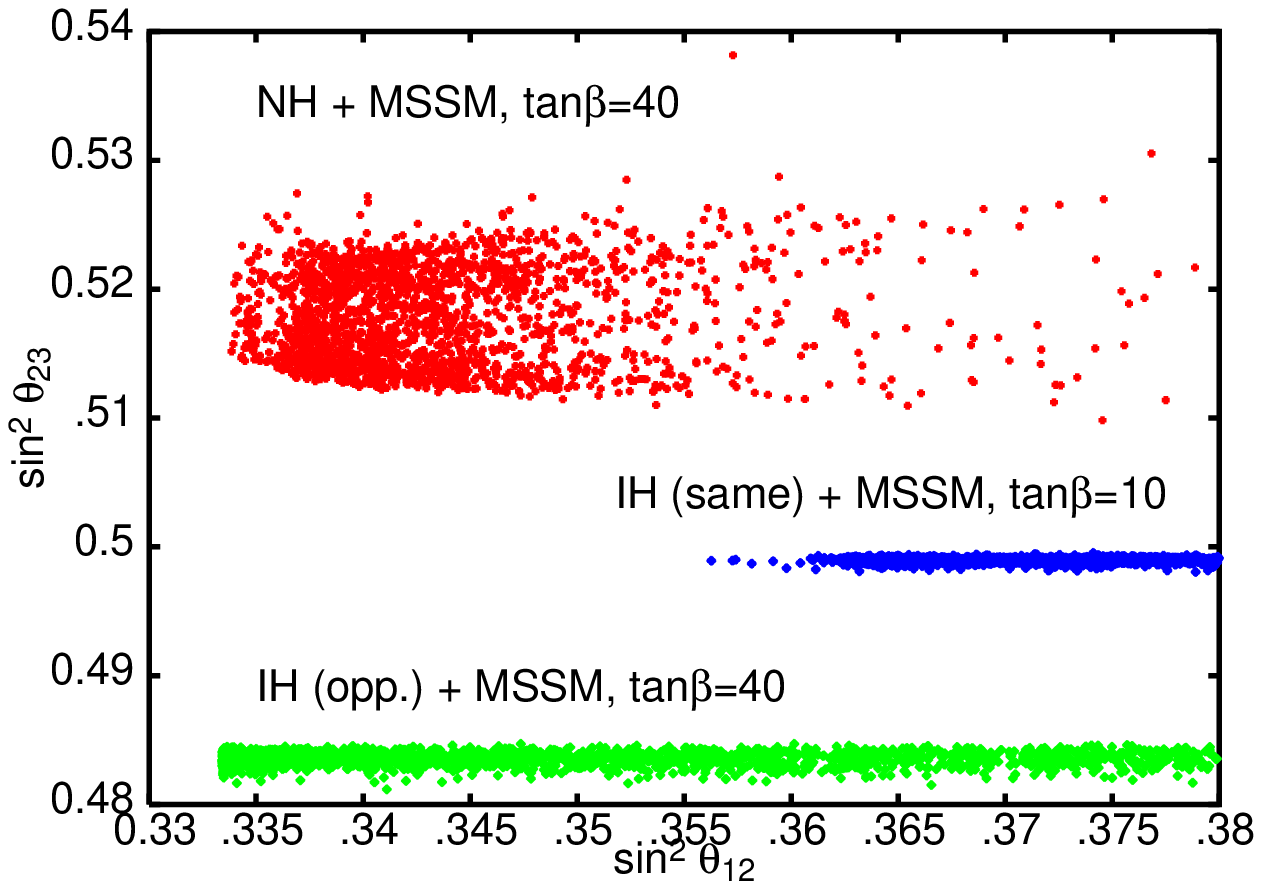,width=12cm,height=8cm}
\caption{\label{fig:RG}Scatter plot of 
$\sin^2 \theta_{12}$ against $\sin^2 \theta_{23}$ 
generated by radiative corrections for different 
situations resulting from Eq.\ (\ref{eq:mnuRG}). 
The parameters $A,B,D$ are taken complex. 
``NH'' means that $|D| \gg |A|, |B|$, ``IH (opp.)'' 
$|A| \simeq |B|/2 \gg |D|$ and ``IH (same)'' $|A| \gg |B| ,|D|$.}
\end{center}
\vspace{-3cm}
\hspace{-3cm}
\epsfig{file=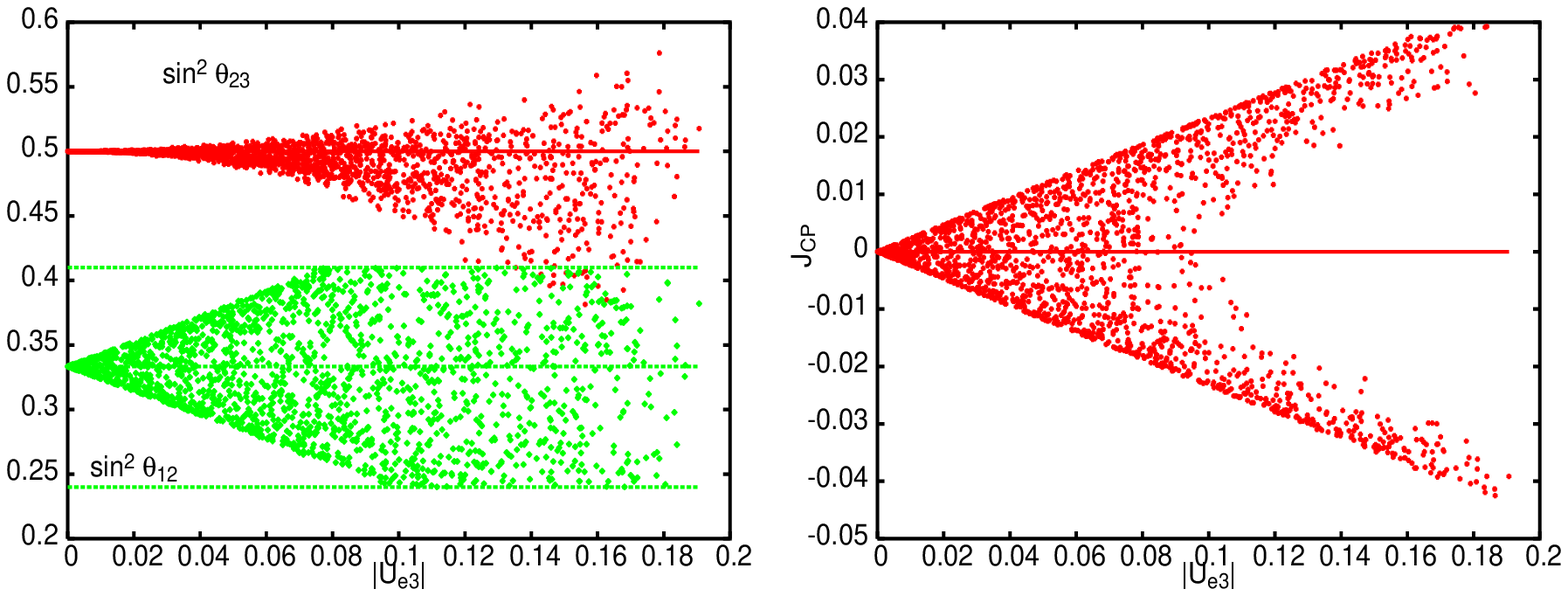,width=22cm,height=29cm}\vspace{-17cm}
\caption{\label{fig:mix}
Scatter plot of $|U_{e3}|$ against $\sin^2 \theta_{12}$ and 
$\sin^2 \theta_{23}$ and against $J_{CP}$ for a ``CKM--like'' matrix 
$U_\ell$. Note that the ranges of $\sin^2 \theta_{12}$ and 
$\sin^2 \theta_{23}$ can overlap.}
\end{figure}

Current neutrino data seems to favor tribimaximal mixing as a very valid 
scenario. Since its predictions include $U_{e3}=0$ and $\theta_{23}=\pi/4$, 
one is interested in breaking scenarios. To get a feeling for the implied 
values of the neutrino mixing angles in such a situation, 
two possibilities have been analyzed 
in this letter. First, the elements of the mass matrix were perturbed and 
second, corrections to the tribimaximal mixing matrix 
from the charged lepton sector were taken into account. 
In the literature one typically focuses on deviations from 
$U_{e3}=0$ and $\theta_{23}=\pi/4$, here we stress that also 
deviations from a zeroth order value of $\theta_{12}$ might 
discriminate between different possible scenarios. 
We summarize the results in Table \ref{tab:sum}. 
Since current data is already very close to tribimaximal mixing, there 
is little room for deviations. Indeed, 
as can be seen from the Table and Figs.\ \ref{fig:NH}, \ref{fig:IH} 
and \ref{fig:RG}, the predicted values of $|U_{e3}|$ 
in case of perturbing the mass matrix are rather small, 
typically below 0.05 for the breaking parameter $\epsilon $ below 0.2.  
This will render them unobservable for the next rounds of experiments. 
Deviations from maximal atmospheric mixing can be up to order 
10 $\%$, which is a testable regime. Only a few cases result in a deviation 
of $\sin^2 \theta_{12}$ from 1/3 of order 10 $\%$.

\noindent  
If the mixing matrix is perturbed by contributions from the 
charged lepton sector one can in principle generate sizable  
deviations for all observables. The interesting case of 
$U_\ell = U_{\rm CKM}$ corresponds to sizable 
$|U_{e3}| \simeq \sin \theta_C /\sqrt{2}$ and large to maximal 
$CP$ violation, but implies only small 
deviations from $\sin^2 \theta_{23}=1/2$ and $\sin^2 \theta_{12}=1/3$. 

\noindent 
One summarizing statement of this work might be the following: 
if $U_{e3}$ deviates sizably from zero, i.e.,  
$U_{e3} \gs 0.1$, then this cannot be achieved by simple 
perturbations of a "tribimaximal" mass matrix. 
Instead, charged lepton contributions are required, which we showed  
to come in this case together with large $CP$ violation.

\vspace{0.5cm}
\begin{center}
{\bf Acknowledgments}
\end{center}
We thank S.~Goswami and R.~Mohapatra for helpful discussions. 
This work was supported by the ``Deutsche Forschungsgemeinschaft'' in the 
``Sonderforschungsbereich 375 f\"ur Astroteilchenphysik'' (F.P.\ and W.R.)
and under project number RO--2516/3--1 (W.R.).

\pagestyle{empty}\vspace{-1cm}
\begin{center}\vspace{-1cm}
\begin{table}[ht]\hspace{-1.8cm}
\begin{tabular}{|c|c|c|}\hline 
 & \multicolumn{2}{c|}{\raisebox{-0.5ex}[0cm][0cm]{Results}} \\[0.1cm]  \cline{2-3}
 \raisebox{1.5ex}[0cm][0cm]{Perturbation}      & \raisebox{-0.7ex}[0cm][0cm]{NH}       & \raisebox{-0.4ex}[0cm][0cm]{IH ($|A| \simeq |B/2|$)} \\[0.1cm] \hline \hline&&\\ 
$\left(
\bad 
A & B & -B(1 + \epsilon) \\[0.2cm]
\cdot & \frac{1}{2} (A + B + D) & \frac{1}{2} (D - A - B)\\[0.2cm]
\cdot & \cdot & \frac{1}{2} (A + B + D)
\ea 
\right)$ 
& 
$\ba 
|U_{e3}| = {\cal O}(\epsilon \sqrt{R}) \\[0.2cm]
\sin^2 \theta_{23} \simeq \frac{1}{2} + {\cal O}(\epsilon^2) \\[0.2cm]
\sin^2 \theta_{12} \simeq \frac{1}{3} ( 1 + \frac{|U_{e3}|}{\sqrt{R}} ) 
\ea $ 
& 
$\ba 
|U_{e3}| \simeq \frac{\epsilon}{4 \sqrt{2}}  \\[0.2cm]
\sin^2 \theta_{23} \simeq \frac{1}{2} (1 + \epsilon ) \\[0.2cm]
\sin^2 \theta_{12} \simeq \frac{1}{3} + \frac{8\sqrt{2}}{27} |U_{e3}|
\ea $ 
\\ &&\\\hline &&\\
$\left(
\bad 
A & B & -B \\[0.2cm]
\cdot & \frac{1}{2} (A + B + D)(1 + \epsilon) 
& \frac{1}{2} (D - A - B)\\[0.2cm]
\cdot & \cdot & \frac{1}{2} (A + B + D)
\ea 
\right)$ 
& 
$\ba 
|U_{e3}| = {\cal O}(\epsilon \sqrt{R}) \\[0.2cm]
\sin^2 \theta_{23} \simeq \frac{1}{2} (1 + \epsilon/2) \\[0.2cm]
\tan 2 \theta_{12} \simeq \frac{2\sqrt{2}}{1 + {\cal O}(\epsilon/\sqrt{R})} 
\ea $ 
& 
$\ba 
|U_{e3}| \simeq \frac{\epsilon}{8\sqrt{2}}\\[0.2cm]
\sin^2 \theta_{23} = \frac{1}{2} ( 1 + |U_{e3}|) \\[0.2cm]
\sin^2 \theta_{12} \simeq \frac{1}{3} (1- {\cal O}(\epsilon) ) 
\ea $ 
\\ &&\\\hline &&\\
$\left(
\bad 
A(1 + \epsilon)  & B & -B \\[0.2cm]
\cdot & \frac{1}{2} (A + B + D)
& \frac{1}{2} (D - A - B)\\[0.2cm]
\cdot & \cdot & \frac{1}{2} (A + B + D)
\ea 
\right)$ 
& 
$\ba 
|U_{e3}| = 0 \\[0.2cm]
\sin^2 \theta_{23} = \frac{1}{2} \\[0.2cm]
\sin^2 \theta_{12} \simeq \frac{1}{3} + \frac{4}{27} \, \epsilon  
\ea $ 
& 
$\ba 
|U_{e3}| = 0 \\[0.2cm]
\sin^2 \theta_{23} = \frac{1}{2} \\[0.2cm]
\sin^2 \theta_{12} \simeq \frac{1}{3} - \frac{2}{27} \, \epsilon  
\ea $ 
\\ &&\\\hline &&\\
$\left(
\bad 
A  & B & -B \\[0.2cm]
\cdot & \frac{1}{2} (A + B + D)
& \frac{1}{2} (D - A - B)(1 + \epsilon)\\[0.2cm]
\cdot & \cdot & \frac{1}{2} (A + B + D)
\ea 
\right)$ 
& 
$\ba 
|U_{e3}| = 0 \\[0.2cm]
\sin^2 \theta_{23} = \frac{1}{2} \\[0.2cm]
\sin^2 \theta_{12} \simeq \frac{1}{3} - \frac{2}{27} \, \epsilon/\sqrt{R}  
\ea $ 
& 
$\ba 
|U_{e3}| = 0 \\[0.2cm]
\sin^2 \theta_{23} = \frac{1}{2} \\[0.2cm]
\sin^2 \theta_{12} \simeq \frac{1}{3} - \frac{1}{27} \, \epsilon  
\ea $ 
\\ &&\\\hline \hline &&\\
 $\left(
\bad 
A & B & -B(1 + \epsilon_{\text{\tiny MSSM}}) \\[0.2cm]
\cdot & \frac{1}{2} (A + B + D) 
& \frac{1}{2} (D - A - B)(1 + \epsilon_{\text{\tiny MSSM}})\\[0.2cm]
\cdot & \cdot & \frac{1}{2} (A + B + D)(1 + 2 \epsilon_{\text{\tiny MSSM}})
\ea 
\right)$
& 
$\ba 
|U_{e3}| = {\cal O}(\epsilon_{\text{\tiny MSSM}} \sqrt{R}) \\[0.2cm]
\sin^2 \theta_{23} \simeq \frac{1}{2} (1 - \epsilon_{\text{\tiny MSSM}}) \\[0.2cm]
\sin^2 \theta_{12} \simeq \frac{1}{3} - \frac{2}{9} \epsilon_{\text{\tiny MSSM}}  
\ea $ 
& 
$\ba 
|U_{e3}| =  {\cal O}(\epsilon_{\text{\tiny MSSM}}^2) \\[0.2cm]
\sin^2 \theta_{23} \simeq \frac{1}{2} (1 + \epsilon_{\text{\tiny MSSM}} ) \\[0.2cm]
\sin^2 \theta_{12} \sim \frac{1}{3}(1 - \epsilon_{\text{\tiny MSSM}}) 
\ea $ 
\\ &&\\\hline \hline \hline 
$U_\ell^\dagger \, U_\nu $ with $U_\ell$ ``CKM--like''& 
\multicolumn{2}{c|}{$\ba\\ 
|U_{e3}| = \frac{\lambda}{\sqrt{2}}~~,~~
J_{CP} \simeq \frac{1}{6} s_\phi \, \lambda \\[0.2cm]
\sin^2 \theta_{23} \simeq \frac{1}{2} + \frac{1}{4} 
\left( c_{2 \phi} + s_{2 \phi} - 2 - 4A \, c_{\omega - \phi} 
\right)\, \lambda^2  \\[0.2cm]
\sin^2 \theta_{12} \simeq \frac{1}{3} - \frac{2\sqrt{2}}{3}|U_{e3}|  
\\[0.5cm] \ea $ }
\\ \hline
$U_{\rm CKM}^\dagger \, U_\nu $ & 
\multicolumn{2}{c|}{$\ba \\
|U_{e3}| = \frac{\sin \theta_C}{\sqrt{2}} ~~,~~
\simeq\mbox{maximal } CP \mbox{ violation}\\[0.2cm]
\sin^2 \theta_{23} \simeq \frac{1}{2}(1 + {\cal O}(|U_{e3}|^2) \\[0.2cm]
\sin^2 \theta_{12} \simeq \frac{1}{3} (1 - |U_{e3}|^2)
\\[0.5cm] \ea $ } \\ \hline \hline 
\end{tabular}
\caption{\label{tab:sum}Typical predictions of the perturbation scenarios 
under discussion.}
\end{table}
\end{center}

\begin{thebibliography}{99} 

\bibitem{S}S.~T.~Petcov,
  Nucl.\ Phys.\ Proc.\ Suppl.\  {\bf 143}, 159 (2005), hep-ph/0412410.




\bibitem{PMNS}B. Pontecorvo, Zh. Eksp. Teor. Fiz.\ {\bf 33}, 549 (1957) 
and {\bf 34}, 247 (1958); 
Z. Maki, M. Nakagawa and S. Sakata, Prog. Theor. Phys.\ {\bf 28}, 870 (1962).

\bibitem{BHP80}
S.~M.~Bilenky, J.~Hosek and S.~T.~Petcov,
Phys.\ Lett.\ B {\bf 94}, 495 (1980); 
M.~Doi {\it et al.},  
Phys.\ Lett.\ B {\bf 102}, 323 (1981); 
J.~Schechter and J.~W.~F.~Valle, Phys.\ Rev.\ D {\bf 23}, 1666 (1981). 

\bibitem{valle}M.~Maltoni {\it et al.},  
New J.\ Phys.\  {\bf 6}, 122 (2004), 
hep-ph/0405172v4.

\bibitem{new_SC}The values for $\sin^2 \theta_{12}$ and $|U_{e3}|^2$ 
have been obtained by A.~Bandyopadhyay, 
S.~Choubey and S.~Goswami by updating the analysis from 
A.~Bandyopadhyay {\it et al.}, 
  Phys.\ Lett.\ B {\bf 608}, 115 (2005) 
with the latest SNO results. 




\bibitem{revs}For recent reviews see, e.g., 
S.~F.~King,
  Rept.\ Prog.\ Phys.\  {\bf 67}, 107 (2004); 
G.~Altarelli and F.~Feruglio,
  New J.\ Phys.\  {\bf 6}, 106 (2004); 
R.~N.~Mohapatra {\it et al.},
hep-ph/0412099. 

\bibitem{tri}
P.~F.~Harrison, D.~H.~Perkins and W.~G.~Scott,
 Phys.\ Lett.\ B {\bf 458}, 79 (1999); 
  Phys.\ Lett.\ B {\bf 530}, 167 (2002); 
P.~F.~Harrison and W.~G.~Scott,
  Phys.\ Lett.\ B {\bf 535}, 163 (2002);
Phys.\ Lett.\ B {\bf 557}, 76 (2003); 
  X.~G.~He and A.~Zee,
  Phys.\ Lett.\ B {\bf 560}, 87 (2003); 
E.~Ma,
  Phys.\ Rev.\ Lett.\  {\bf 90}, 221802 (2003); 
  Phys.\ Lett.\ B {\bf 583}, 157 (2004); 
  C.~I.~Low and R.~R.~Volkas,
  Phys.\ Rev.\ D {\bf 68}, 033007 (2003); 
S.~H.~Chang, S.~K.~Kang and K.~Siyeon,
  Phys.\ Lett.\ B {\bf 597}, 78 (2004); 
E.~Ma, 
Phys.\ Rev.\ D {\bf 70}, 031901 (2004); 
F.~Caravaglios and S.~Morisi,
  hep-ph/0503234; 
G.~Altarelli, F.~Ferruglio, hep-ph/0504165;  
E.~Ma, hep-ph/0505209; 
I.~de Medeiros Varzielas and G.~G.~Ross, 
hep-ph/0507176; 
K.~S.~Babu and X.~G.~He, hep-ph/0507217. 


\bibitem{devtri0}
A.~Zee,
  Phys.\ Rev.\ D {\bf 68}, 093002 (2003); 
N.~Li and B.~Q.~Ma,
  Phys.\ Rev.\ D {\bf 71}, 017302 (2005). 

\bibitem{devtri11}
Z.~Z.~Xing,
  Phys.\ Lett.\ B {\bf 533}, 85 (2002).

\bibitem{devtri12}
S.~F.~King, hep-ph/0506297. 


\bibitem{tri1}
L.~Wolfenstein,
  Phys.\ Rev.\ D {\bf 18}, 958 (1978).



\bibitem{sno2}
  B.~Aharmim {\it et al.}  [SNO Collaboration],
  nucl-ex/0502021.

\bibitem{fut}
S.~Choubey and W.~Rodejohann,
  hep-ph/0506102; 
S.~Goswami, talk given at XXII International Symposium on 
Lepton--Photon Interactions at High Energy "Lepton/Photon 05", Uppsala, 
Sweden, July 2005; 
{\tt http://lp2005.tsl.uu.se/~lp2005/LP2005/programme/index.htm}

\bibitem{chef}P.~Huber {\it et al.}, 
  Phys.\ Rev.\ D {\bf 70}, 073014 (2004). 

\bibitem{pp}J.~N.~Bahcall and C.~Pena-Garay,
JHEP {\bf 0311}, 004 (2003). 

\bibitem{SPMIN}A.~Bandyopadhyay {\it et al.}, 
hep-ph/0410283.

\bibitem{bimax}F.~Vissani, hep-ph/9708483; 
V.~D.~Barger, S.~Pakvasa, T.~J.~Weiler and K.~Whisnant, 
Phys.\ Lett.\ B {\bf 437}, 107 (1998); 
A.~J.~Baltz, A.~S.~Goldhaber and M.~Goldhaber, 
Phys.\ Rev.\ Lett.\  {\bf 81}, 5730 (1998); 
H.~Georgi and S.~L.~Glashow,
Phys.\ Rev.\ D {\bf 61}, 097301 (2000); 
I.~Stancu and D.~V.~Ahluwalia,
Phys.\ Lett.\ B {\bf 460}, 431 (1999). 


\bibitem{mutau}C.~S.~Lam,
  Phys.\ Lett.\ B {\bf 507}, 214 (2001); 
  W.~Grimus and L.~Lavoura,
  JHEP {\bf 0107}, 045 (2001); 
  Eur.\ Phys.\ J.\ C {\bf 28}, 123 (2003); 
J.\ Phys.\ G {\bf 30}, 1073 (2004); 
  T.~Kitabayashi and M.~Yasue,
  Phys.\ Lett.\ B {\bf 524}, 308 (2002); 
  I.~Aizawa {\it et al.}, 
  Phys.\ Rev.\ D {\bf 70}, 015011 (2004); 
  W.~Grimus {\it et al.},  
  JHEP {\bf 0407}, 078 (2004); 
R.~N.~Mohapatra, S.~Nasri and H.~B.~Yu, 
Phys.\ Lett.\ B {\bf 615}, 231 (2005); 
  R.~N.~Mohapatra and S.~Nasri,
  Phys.\ Rev.\ D {\bf 71}, 033001 (2005); 
C.~S.~Lam,
  Phys.\ Rev.\ D {\bf 71}, 093001 (2005).

\bibitem{ma}
P.~F.~Harrison and W.~G.~Scott,
  Phys.\ Lett.\ B {\bf 547}, 219 (2002); 
E.~Ma,
  Phys.\ Rev.\ D {\bf 66}, 117301 (2002); 
I.~Aizawa, T.~Kitabayashi and M.~Yasue, hep-ph/0504172; 
T.~Kitabayashi and M.~Yasue, hep-ph/0504212. 


\bibitem{Grimus:2004cc}
  W.~Grimus {\it et al.},  
  Nucl.\ Phys.\ B {\bf 713}, 151 (2005). 



\bibitem{lmlt}
P.~Binetruy {\it et al.}, 
Nucl.\ Phys.\ B {\bf 496}, 3 (1997); 
N.~F.~Bell and R.~R.~Volkas,
Phys.\ Rev.\ D {\bf 63}, 013006 (2001);  
S.~Choubey and W.~Rodejohann,
Eur.\ Phys.\ J.\ C {\bf 40}, 259 (2005). 


\bibitem{A4}K.~S.~Babu, E.~Ma and J.~W.~F.~Valle, 
Phys.\ Lett.\ B {\bf 552}, 207 (2003); 
M.~Hirsch {\it et al.},  
Phys.\ Rev.\ D {\bf 69}, 093006 (2004).


\bibitem{rabimutau}R.~N.~Mohapatra,
  JHEP {\bf 0410}, 027 (2004). 

\bibitem{RGE}See for instance 
S.~Antusch {\it et al.}, 
JHEP {\bf 0503}, 024 (2005) 
and references therein. 

\bibitem{devbimax}
M.~Jezabek and Y.~Sumino,
  Phys.\ Lett.\ B {\bf 457}, 139 (1999); 
Z.~Z.~Xing,
  Phys.\ Rev.\ D {\bf 64}, 093013 (2001); 
C.~Giunti and M.~Tanimoto,
  Phys.\ Rev.\ D {\bf 66}, 053013 (2002); 
  Phys.\ Rev.\ D {\bf 66}, 113006 (2002); 
W.~Rodejohann,
  Phys.\ Rev.\ D {\bf 69}, 033005 (2004); 
G.~Altarelli, F.~Feruglio and I.~Masina,
  Nucl.\ Phys.\ B {\bf 689}, 157 (2004); 
A.~Romanino,
  Phys.\ Rev.\ D {\bf 70}, 013003 (2004). 

\bibitem{QLC}
M.~Raidal,
  Phys.\ Rev.\ Lett.\  {\bf 93}, 161801 (2004); 
H.~Minakata and A.~Y.~Smirnov,
  Phys.\ Rev.\ D {\bf 70}, 073009 (2004); 
P.~H.~Frampton and R.~N.~Mohapatra,
  JHEP {\bf 0501}, 025 (2005); 
J.~Ferrandis and S.~Pakvasa,
  Phys.\ Rev.\ D {\bf 71}, 033004 (2005); 
N.~Li and B.~Q.~Ma,
Phys.\ Rev.\ D {\bf 71}, 097301 (2005); hep-ph/0504161; 
S.~K.~Kang, C.~S.~Kim and J.~Lee,
  hep-ph/0501029; 
K.~Cheung {\it et al.}, 
hep-ph/0503122; 
Z.~Z.~Xing,
  hep-ph/0503200; 
A.~Datta, L.~Everett and P.~Ramond,
  hep-ph/0503222; 
S.~Antusch, S.~F.~King and R.~N.~Mohapatra,
  hep-ph/0504007; 
H.~Minakata,
  hep-ph/0505262; 
T.~Ohlsson,
  hep-ph/0506094.

\bibitem{FPR}
P.~H.~Frampton, S.~T.~Petcov and W.~Rodejohann,
  Nucl.\ Phys.\ B {\bf 687}, 31 (2004); 
  S.~T.~Petcov and W.~Rodejohann,
  Phys.\ Rev.\ D {\bf 71}, 073002 (2005). 




\bibitem{CJ85}C.~Jarlskog, 
Z.\ Phys.\ C {\bf 29}, 491 (1985); 
Phys.\ Rev.\ D {\bf 35}, 1685 (1987).

\bibitem{PKSP3nu88}
P.~I.~Krastev and S.~T.~Petcov, 
Phys.\ Lett.\ B {\bf 205} (1988) 84.


\bibitem{JMaj87} 
J.~F.~Nieves and P.~B.~Pal, 
Phys.\ Rev.\ D {\bf 36}, 315 (1987); 
Phys.\ Rev.\ D {\bf 64}, 076005 (2001); 
J.~A.~Aguilar-Saavedra and G.~C.~Branco, 
Phys.\ Rev.\ D {\bf 62}, 096009 (2000).  


\end{thebibliography}
\end{document}